\documentclass[aps,a4paper,amsmath,showpacs,final,twocolumn,floatfix]{revtex4}
\usepackage{graphicx}
\usepackage{amsmath}

\begin{document}

\title{Atom lithography with two-dimensional optical masks}

\author{S.J.H.~Petra}
\author{K.A.H.~van~Leeuwen}
\author{L.~Feenstra}
\altaffiliation{\emph{Present address}: Physikalisches Institut, Universit{\"a}t Hei\-del\-berg, Philosophenweg 12, 69120 Heidelberg, Germany}
\author{W.~Hogervorst}
\author{W.~Vassen}

\affiliation{Atomic and Laser Physics Group, Laser Centre Vrije Universiteit, De Boelelaan 1081, 1081 HV Amsterdam, The Netherlands}


\begin{abstract}
With a two-dimensional (2D) optical mask at $\lambda=1083$~nm, nanoscale patterns are created for the first time in an atom lithography process using metastable helium atoms.
The internal energy of the atoms is used to locally damage a hydrofobic resist layer, which is removed in a wet etching process.
Experiments have been performed with several polarizations for the optical mask, resulting in different intensity patterns, and corresponding nanoscale structures.
The results for a linear polarized light field show an array of holes with a diameter of 260~nm, in agreement with a computed pattern.
With a circularly polarized light field a line pattern is observed with a spacing of $\lambda/\sqrt{2} = 766$~nm.
Simulations taking into account many possible experimental imperfections can not explain this pattern.
\end{abstract}

\maketitle

\section{Introduction}
\label{2dintro}
Over the last ten years atomic nanofabrication with optical masks has been demonstrated in numerous experiments.
With a direct deposition technique, 1D structures have been grown with sodium \cite{timp92}, chromium \cite{mccl93,drod97,sun01,jurd02}, aluminum \cite{mcgo95}, ytterbium \cite{ohmu03} and iron \cite{leeu04}.
Similarly, in a two-step lithography process, experiments have been performed with cesium \cite{liso97}, metastable argon \cite{john98}, metastable neon \cite{enge99}, and metastable helium \cite{petr04}.
In a lithography process, a metal film is covered with an organic resist layer, which is selectively damaged by the studied atoms.
Next, the damaged molecules and their underlying metal layer are removed with a cyanide etching solution.
Although the lithography procedure requires an additional step, it has the advantage that the pattern formation does not suffer from atomic diffusion in the growth process \cite{slig04}, and the structures have no pedestal.

So far, experiments with a 2D optical mask have been performed by deposition of chromium atoms \cite{brad99,schu00}.
Chromium is advantageous because a short wavelength can be used ($\lambda = 426$~nm) implying small structure sizes of 70 -- 80~nm.
Disadvantage is the long exposure time and the inevitable pedestal in the structures.
Also, atom lithography with cesium atoms and a holographic light mask has been reported, resulting in a line pattern with a periodicity of 426~nm and a perpendicularly superposed line pattern with a $23.4~\mu$m spacing \cite{mutz02}.
We report on experiments with a 2D optical mask using metastable helium atoms in a lithography procedure.
The metastable helium atom has a high internal energy (20~eV), and the dose of atoms required to affect a single resist molecule is therefore low \cite{nowa96}.
This shortens the exposure time of the samples to about eight minutes \cite{petr04}.

The optical mask is generated by interference of laser beams that propagate perpendicularly to an atomic beam which is directed towards the sample.
A force is exerted on the atom by the interaction of the induced atomic dipole moment with the electric field of the standing light wave.
For a two-level atom, the optical potential of the standing-wave light field can be written as \cite{ashk78}:
\begin{equation}
\label{udip}
U(\vec{r}) = \frac{\hbar \Delta}{2} \ln\left(1+\frac{I(\vec{r})} {I_{\textrm{sat}}} \frac{\Gamma^2}{\Gamma^2 + 4 \Delta^2}\right),
\end{equation}
where $\Delta$ is the detuning of the light from atomic resonance, $I(\vec{r})$ is the intensity of the light as a function of the position $\vec{r}=(x,y,z)$, $I_{\textrm{sat}}$ is the saturation intensity of the optical transition, and $\Gamma$ is the natural linewidth of the atomic transition.
The dipole force experienced by the atom, corresponding to this optical potential, is given by:
\begin{equation}
\label{fdip}
\vec{F}(\vec{r}) = -\vec{\nabla} U(\vec{r}).
\end{equation}

\begin{figure}[t]
\includegraphics[width=\columnwidth,keepaspectratio]{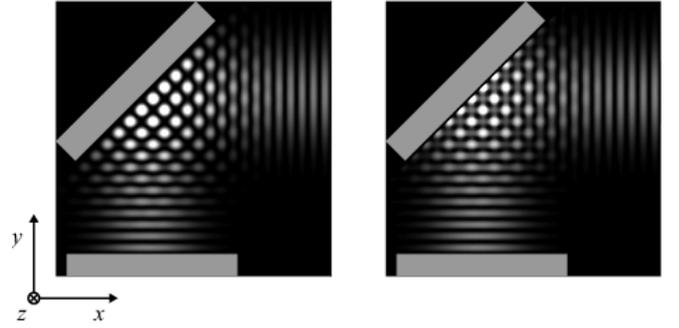}
\caption{\label{intensitypics}
Schematic view of the 2D intensity profiles for linearly polarized light, perpendicular to the plane of incidence (left-hand image) and circularly polarized light (right-hand image).
The mirrors are indicated by the gray rectangles.
The ratio of the wavelength of the light and the waist of the laser beam is not to scale.}
\end{figure}

The formation of a 2D optical mask is shown schematically in Figure~\ref{intensitypics}.
The incident light beam, traveling from right to left in the $x$-direction, is reflected at $45^{\circ}$ downward in the minus $y$-direction and retro-reflected by the lower mirror.
In this mirror configuration a 2D interference pattern appears just in front of the upper mirror.
The atomic beam propagates in the $z$-direction through the light field.
The samples are placed in the light field in front of this upper mirror.
The two images of Figure~\ref{intensitypics} show the intensity profile of the optical mask for the two different polarizations used in the experiments.
The left-hand image presents the pattern formed by a linearly polarized light beam, polarized perpendicular to the plane of incidence, \emph{i.e.} in the $z$-direction.
The right-hand image shows the intensity profile of a circularly polarized light beam.

\section{Experimental setup}
\label{2dsetup}
\begin{figure}[t]
\includegraphics[width=\columnwidth,keepaspectratio]{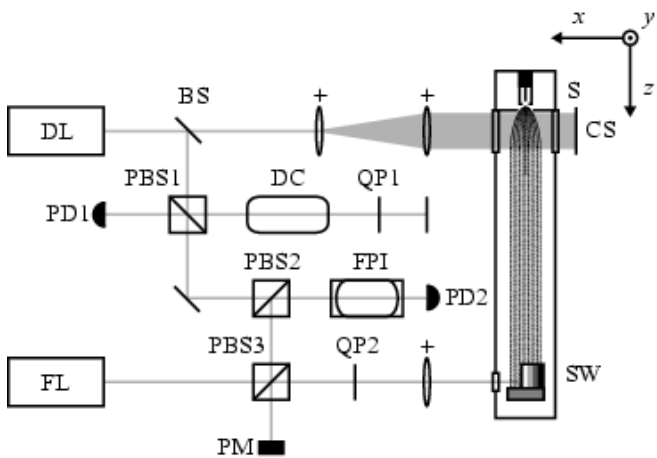}
\caption{\label{2dexpsetup}
Schematic of the setup.
Metastable helium atoms are produced in DC discharge source S and travel in the $z$-direction toward the sample.
In the collimation section CS, laser beams of diode laser DL allow for collimation of the atomic beam in both $x$- and $y$-direction.
Part of the diode laser output (4\%) is split of by beam splitter BS for spectroscopy.
After reflection by polarizing beam splitter PBS1 and double-passing discharge cell DC and quarter-wave plate QP1, the light is detected by photodiode PD1\@.
The light transmitted by PBS1 also passes PBS2 and is analyzed with Fabry-Perot interferometer FPI and photodiode PD2\@.
The optical standing-wave mask is made by the fiber laser FL\@.
A small part of its beam reflected by PBS3 and also analyzed by FPI\@.
For the standing wave the beam is first reflected downward in the $y$-direction and retro-reflected by the mirror set SW, which is incorporated in the sample holder inside the vacuum chamber.
After double-passing QP2, the light of the standing wave is reflected by PBS3 and monitored with power meter PM\@.}
\end{figure}

The experimental setup is shown in Figure~\ref{2dexpsetup} and is similar to the setup we described in a previous paper \cite{petr04}.
The helium atoms are transferred to the metastable $2~^3\textrm{S}_1$ state in a DC helium discharge.
The atomic beam is collimated in two dimensions using a curved wavefront technique with converging laser beams \cite{aspe90,rooi96}.
After the 2D-collimation the transverse velocity spread of the atomic beam, defined as the full width at $1/\textrm{e}^2$ height, is decreased to 3~m/s and the centerline beam intensity is increased sixfold to $1.1 \times 10^{10}~\textrm{s}^{-1}~\textrm{mm}^{-2}$ on the sample.
The laser light for the collimation beams is provided by a 20~mW distributed Bragg reflector diode laser that operates at a wavelength of 1083~nm.
With Doppler-free saturation spectroscopy this laser is actively stabilized to the $2~^3\textrm{S}_1 \rightarrow 2~^3\textrm{P}_2$ transition of the helium atom.

The light for the 2D optical mask is obtained from a fiber laser also operating at a wavelength of 1083~nm and delivering a power of 800~mW.
The frequency of this light is 375~MHz blue detuned with respect to the $2~^3\textrm{S}_1 \rightarrow 2~^3\textrm{P}_2$ transition.
The atoms are therefore repelled to the intensity nodes of the standing-wave light field.
The frequency stability of the fiber laser is monitored with respect to the locked diode laser with a Fabry-Perot interferometer (see Figure~\ref{2dexpsetup}).
The laser beam is focused at the sample holder with a waist (defined as radius at $1/\textrm{e}^2$ of the intensity height) of 0.33~mm.
The corresponding Rayleigh length of 32~cm renders the actual focal point uncritical.
The maximum saturation parameter is $I_{\textrm{max}}/I_{\textrm{sat}} = 10^7$, where $I_{\textrm{max}}$ is the maximum intensity at the nodes of the standing wave and $I_{\textrm{sat}} = 0.17$~mW/cm$^2$.
The retro-reflected laser beam overlaps the incident beam path over a distance of 3.3~m, and is then reflected by polarizing beam splitter cube PBS3 and monitored with a power meter.
The standing-wave is aligned with respect to the sample position by first adjusting it such that it is not clipped by the sample and thus 100\% of the light is back-reflected by the standing-wave mirror onto the power meter.
Next, the beam is displaced by adjusting a micrometer translation stage (not indicated in figure~\ref{2dexpsetup}) and clipped by the sample until only 50\% of the light is back-reflected, which indicates that the sample position is at the center of the laser beam.
Final fine tuning is done with the translation stage in order to move the laser beam to the desired position with respect to the sample.
The large distance over which both the incident and retro-reflected beam overlap reduces the misalignment angle of the laser beam to less than 0.3~mrad.
In the case where a linearly polarized optical mask is used, quarter-wave plate QP2 is removed after alignment of the laser beam.

The samples are silicon substrates covered with a 1~nm chromium layer and a 30~nm gold layer.
A hydrofobic resist layer of nonanethiol [CH$_3${(CH$_2$)}$_8$SH] molecules allows to develop the samples in a simple wet-etching process \cite{petr04,nowa96,xia95}.
After depositing a Self-Assembled Monolayer (SAM) of nonanethiol, the samples are exposed to the atomic beam for eight minutes, corresponding to a dose of about one metastable atom per SAM molecule.
In a Penning ionization process, the SAM is locally damaged by the metastable atoms \cite{ozak90}.
After exposure, the samples are etched in a cyanide solution that dissolves the damaged molecules, and the underlying gold film \cite{petr04,nowa96,xia95}.
The etching time is also typically eight minutes.
After etching, the sample topography is analyzed with an Atomic Force Microscope (AFM).

\section{Results}
\label{2dresults}
\begin{figure}[t]
\includegraphics[width=\columnwidth,keepaspectratio]{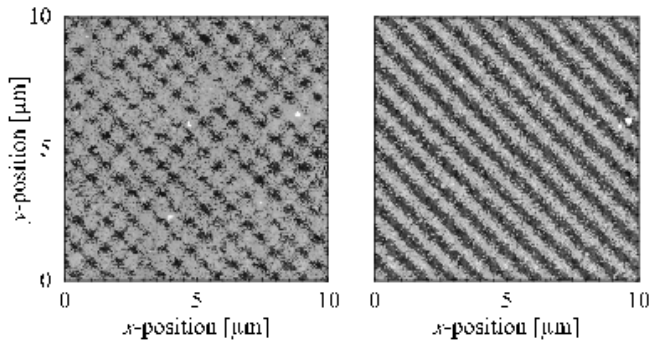}
\caption{\label{2dexpresults}
AFM surface scans of the samples.
The images cover an area of $10~\mu\textrm{m} \times 10~\mu\textrm{m}$.
The left-hand and right-hand images show the results of the linearly polarized and the circularly polarized optical mask respectively.
The separation between both the lines and the dots is $\lambda/\sqrt{2} = 766$~nm.}
\end{figure}

Figure~\ref{2dexpresults} shows two $10~\mu\textrm{m} \times 10~\mu\textrm{m}$ surface scans of the samples taken with the AFM\@.
The structure shown in the left-hand image was created with a linearly polarized optical mask.
The dark regions indicate the positions where the metastable helium atoms have hit the sample surface and where the gold layer has been removed in the etching process.
The average structure height is about 20~nm.
The image shows a lattice of dots with a $\lambda/\sqrt{2} = 766$~nm separation.
The holes have an average Full Width at Half Maximum (FWHM) diameter of 260~nm.
The right-hand image pattern was produced with a circularly polarized optical mask.
The orientation of the lines is perpendicular to the mirror surface, and the line separation is also $\lambda/\sqrt{2}$, with an average diameter of 360~nm.
A close look at the structures shows a graininess typical for all lithography experiments with gold layers \cite{liso97,petr04,nowa96}.
Zooming in on a part of the figure shows that the gold typically consists of approximately 20~nm diameter grains.

\section{Discussion}
\label{2ddiscussion}
\begin{figure}[t]
\includegraphics[width=\columnwidth,keepaspectratio]{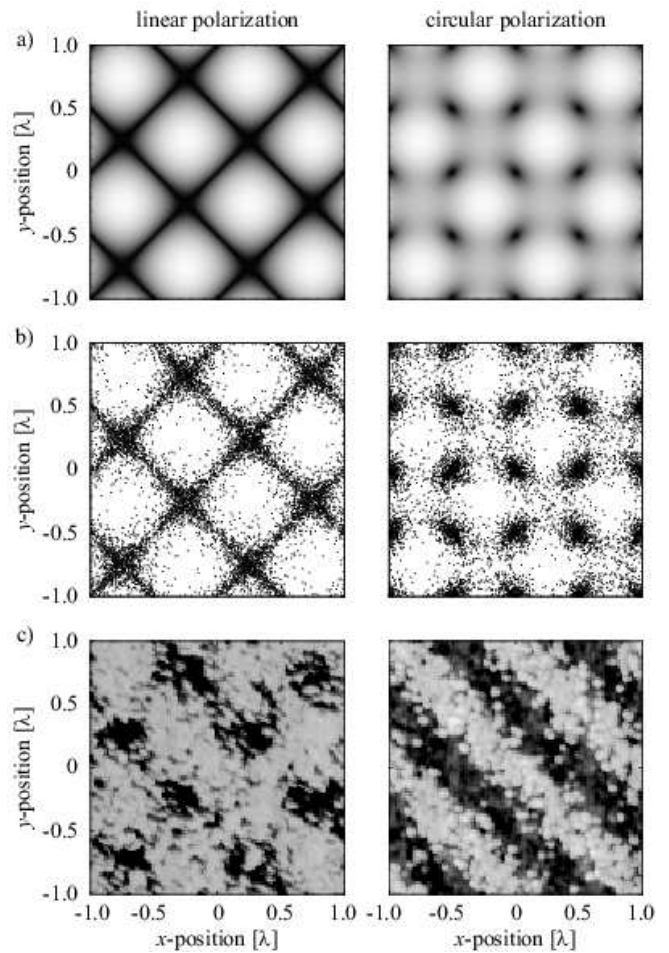}
\caption{\label{2dcalculations}
Calculations of the potential (a), the atomic distributions on the sample (b) and the experimental results (c) for a circularly polarized light field (left-hand images) and linearly polarized light field (right-hand images).
All images cover an area of $2 \lambda \times 2 \lambda$, with $\lambda$ the wavelength of the light.}
\end{figure}

For a better understanding of the created structures, calculations of both the optical potential of the light field and the expected atomic distributions on the samples have been performed.
Due to the magnetic degeneracy of the lower and upper states of the optical transition, the potential of Equation~\ref{udip} is not strictly valid.
In a more accurate approach, the actual (adiabatic) potentials have been calculated by diagonalizing the interaction with the light field.
However, the resulting potentials for the $\textrm{J}=1 \rightarrow \textrm{J}=2$ transition used here correspond closely to the two-level potential of Equation~\ref{udip}.
Hence, for convenience this potential has been used for the calculations.

The potentials of the optical masks corresponding to the intensity profiles of Figure~\ref{intensitypics} are shown in Figure~\ref{2dcalculations}a for linearly polarized light (left-hand image) and circularly polarized light (right-hand image).
The images cover an area of $2 \lambda \times 2 \lambda$, where $\lambda=1083$~nm is the wavelength of the optical transition.
The gray-scale indicates the height of the potential, where the lightest regions correspond to the areas where the potential is the highest.
For both polarizations, numerical simulations have been performed on the atoms that travel through the 2D standing-wave light field.
These calculations were done using the dipole force $\vec{F}(\vec{r})$ from Equation~\ref{fdip} in a three-dimensional model that calculates the atomic trajectories numerically.
In a previous paper we have described this model in detail \cite{petr03}.
The atomic distributions corresponding to the potential plots are shown in Figure~\ref{2dcalculations}b, where every black dot represents the position where an atom has hit the sample.
As an initial distribution, a lattice of $201 \times 201$ atoms is used, distributed homogeneously on a $4 \lambda \times 4 \lambda$ grid.
Both the longitudinal and transverse velocity of every atom is picked randomly from a Gaussian velocity distribution that corresponds to the experimental values.
The calculations, simulations and measurements in Figure~\ref{2dcalculations} were taken at a sample position of $100~\mu$m behind the center of the laser beam, in the downstream direction of the atomic beam.
From Figure~\ref{2dcalculations}a and Figure~\ref{2dcalculations}b it is clear that the atoms are guided to the regions of the lowest potential, \emph{i.e.}, the dark regions of the potential plots.
The calculated structure for linearly polarized light (along the $z$-direction) shows a grid structure of the atoms, with a line separation of $\lambda/\sqrt{2}$.
For the circularly polarized light field, a dotted pattern is formed, where the separation of two successive dots is $\lambda/\sqrt{2}$ at an angle of $45^{\circ}$ with the $x$-axis.

For comparison, the experimental results for both polarizations are shown in Figure~\ref{2dcalculations}c as well.
These images are close-ups of a part ($2 \lambda \times 2\lambda$) of the images shown in Figure~\ref{2dexpresults}.
The images with a linearly polarized light field are in agreement with the simulated atomic distribution shown in Figure~\ref{2dcalculations}b.
The qualitative visual agreement between the patterns is confirmed by a 2D Fourier analysis of the distribution.
The calculated diameter (FWHM) of the dots (along the $x$- and $y$-axis) is 170~nm, which is in agreement with the measured width of 260~nm.
The broadened feature size seen in the experimental results can be attributed to the isotropic etch process, that removes the unprotected gold from all directions.
The lack of visibility of the lines connecting the dots is probably because the dose of metastable atoms was too small to damage the resist layer enough in those specific areas.
A longer exposure time, however, will broaden the dots, similar to the 1D case \cite{petr04}.

The pattern formed with a circularly polarized light field deviates strongly from the calculated structure.
Instead of a lattice of dots, the experimental results show a line pattern orientated perpendicular to the upper stand\-ing-wave mirror surface.
Despite calculations that take into account a possible misalignment of the laser beam, an imperfect polarization state and/or intensity gradients of the beams due to imperfect mirrors, we can not explain the observed pattern.
%Also an extended model that accounts for the light coupling between the different magnetic substates of the metastable helium atom does not describe the experimental results.
Both mirrors for the 2D optical masks have been verified experimentally to have a conservation of polarization better than 97\%.
Thorough examination of the samples with the AFM have shown that the line pattern seen in Figure~\ref{2dexpresults} is clearly visible in a large area of $300~\mu\textrm{m} \times 300~\mu\textrm{m}$.
Outside this area the 1D standing wave dominates the interference pattern, resulting in line structures with a separation of $\lambda/2$, and indicating that the 2D structure is produced at the center of the 2D light field.
This is confirmed by the structure direction and spacing ($\lambda/\sqrt{2}$), that can only be explained from the atom having experienced the force of a 2D optical mask.
Therefore the observed structures with a circularly polarized light field remain unexplained.

\section{Conclusion}
\label{2dconclusion}
We have demonstrated atom lithography with metastable helium using 2D optical masks.
Calculations can only partly explain the observed structures.
With a linearly polarized light field a lattice of holes with a diameter (FWHM) of 260~nm is created.
The structures obtained with a circularly polarized optical mask show lines in a direction perpendicular to the mirror surface with a FWHM of 360~nm.
The separation between both the lines and the holes is $\lambda/\sqrt{2} = 766$~nm.
The structure size that may be written with 1083~nm light in a 2D mask is determined by the wavelength of the light.
Structures with a smaller periodicity and size are feasible when shorter wavelength light is used.
In this respect the $2~^3\textrm{S}_1 \rightarrow 3~^3\textrm{P}_2$ transition is promising.
Light at the $\lambda = 389$~nm wavelength can be generated efficiently by frequency doubling of a Ti:Sapphire laser \cite{koel03}.
The recent advent of diode lasers at this wavelength is promising as well.
However, the problem with the 20~nm graininess of the structures remains, and has to be solved when structures smaller then 20~nm are aimed for.

\begin{acknowledgments}
We are grateful to E.W.J.M.\ van der Drift (DIMES Nanofacility, The Netherlands) for providing the silicon wafers.
Financial support from the Foundation for Fundamental Research on Matter (FOM) is gratefully acknowledged.
\end{acknowledgments}

\bibliographystyle{apsrev}

\end{document}